\documentclass[aps,prd,preprint]{revtex4}
\usepackage{graphicx}

\newcommand{\gsim}{\lower.7ex\hbox{$\;\stackrel{\textstyle>}{\sim}\;$}}
\newcommand{\lsim}{\lower.7ex\hbox{$\;\stackrel{\textstyle<}{\sim}\;$}}

\begin{document}

\title{Hard pomeron enhancement of ultrahigh-energy neutrino-nucleon
cross-sections}

\author{A. Z. Gazizov and S. I. Yanush}
\email{gazizov@dragon.bas-net.by}
\author{S. I. Yanush}
\email{yanush@dragon.bas-net.by} \affiliation{B. I. Stepanov
Institute of Physics of the National Academy of Sciences of
Belarus,\\ F. Skariny Ave.\ 68, 220072 Minsk, Belarus}

\begin{abstract}
An unknown small-$x$ behavior of nucleon structure functions gives
appreciable uncertainties to high-energy neutrino-nucleon
cross-sections. We construct structure functions using at small
$x$ Regge inspired description by A.~Donnachie and P.~V.~Landshoff
with \emph{soft} and \emph{hard} pomerons, and employing at
larger $x$ the perturbative \emph{QCD} expressions. The smooth
interpolation between two regimes for each $Q^2$ is provided with
the help of simple polynomial functions. To obtain low-$x$
neutrino-nucleon structure functions $F_{2}^{\nu N, \bar \nu
N}(x,Q^2)$ and singlet part of $F_{3}^{\nu N,\bar \nu N}(x,Q^2)$
from Donnachie-Landshoff function $F_2^{ep}(x,Q^2)$, we use the
$Q^2$-dependent ratios $R_2(Q^2)$ and $R_3(Q^2)$ derived from
perturbative \emph{QCD} calculations. Non-singlet part of $F_3$
at low $x$, which is very small, is taken as power-law
extrapolation of perturbative function at larger $x$. This
procedure gives a full set of smooth neutrino-nucleon structure
functions in the whole range of $x$ and $Q^2$ at interest.

Using these structure functions, we have calculated the
neutrino-nucleon cross-sections and compared them with some other
cross-sections known in literature. Our cross-sections turn out to
be the highest among them at the highest energies, which is
explained by contribution of the \emph{hard} pomeron.

\end{abstract}

\maketitle

%%%%%%%%%%%%%%%%%%%%%%%%%%%%%%%%%%%%%

\section{Introduction}

The interest to neutrino-nucleon cross-sections at very high
energies, up $\sim 10^{21}$~eV, is stimulated by \emph{High Energy
Neutrino Astronomy (HENA)} (for a review see \cite{book,GHS}).
\emph{Ultra High Energy (UHE)} neutrinos can be of accelerator and
non-accelerator origin. In the former case \emph{UHE} protons
accelerated in astrophysical sources produce neutrinos in the
chain of $\pi$- and $K$-decays, when \emph{UHE} protons interact
with ambient gas or with low energy photons. Since from
observations we know that \emph{UHE} protons exist with energies
up to $3\times 10^{20}$~eV \cite{CR_3^10*20}, the maximum energy
of neutrinos is expected up to $\sim 10^{19}$~eV. Astrophysical
accelerators are usually connected with shock waves in \emph{SNe},
\emph{AGNs}, \emph{GRBs} etc, but there could be also some other
mechanisms of acceleration, such as acceleration in the strong
electromagnetic wave and in strong electric field due to unipolar
inductors (see Ref.\cite{book}).

Non-accelerator sources can provide neutrinos even with higher
energies. These sources include production by Topological Defects
(first suggested in Ref.\cite{HiSch}), by decays of superheavy
dark matter particles and by annihilation of superheavy particles.
Topological Defects in many cases become unstable and decompose to
constituent fields, superheavy gauge bosons and Higgs particles,
which then decay to hadrons and neutrinos. There could be the
examples when the constituent superheavy fields are produced at
annihilation (e.g.\ annihilation of monopole-antimonopole
connected by string). Annihilation of dark matter particles (e.g.\
neutralino in \emph{Earth} and \emph{Sun}) gives another source of
high energy neutrino production. The maximum energy of neutrinos
from above-mentioned sources can reach the \emph{GUT} scale.

High energy neutrino radiation from all sources is inevitably
accompanied by other radiations, most notably by high energy gamma
rays. Even in cases when high energy photons are absorbed in the
source, their energy is partly transformed into low energy photon
radiation: $X$-rays and thermal radiation. For sources transparent
for high energy gamma radiation the upper limit on diffuse
neutrino flux is imposed by the cascade electromagnetic radiation
(see Ref.\cite{book}). Colliding with microwave photons, high
energy photons and electrons give rise to electromagnetic cascades
with most of energy being in the observed $100$~MeV -- $10$~GeV
energy range. The energy density of this cascade radiation should
not exceed, according to \emph{EGRET} observations, $\omega_{cas}
\sim (1-2)\times 10^{-6}$ eV/cm$^3$. Introducing the neutrino
energy density for neutrinos with individual energies higher than
$E$, $\omega_{\nu}(>E)$, it is easy to derive the following chain
of inequalities (from left to right):
\begin{equation}
\omega_{cas}>\omega_{\nu}(>E)=%
\frac{4\pi}{c}\int_E^{\infty}EI_{\nu}(E)dE=%
\frac{4\pi}{c}E I_{\nu}(>E). %
\label{cas}
\end{equation}
An upper bound on integral neutrino flux immediately follows from
Eq.~(\ref{cas}):
\begin{equation}
I_{\nu}(>E)<\frac{c}{4\pi}\frac{\omega_{cas}}{E}.%
\label{upbound}
\end{equation}
The latter inequality gives a powerful limit on the possible
diffuse neutrino flux.

\emph{UHE} neutrinos can be detected at the \emph{Earth} due to
their interactions with nucleons in \emph{CC} and \emph{NC}
reactions,
\begin{eqnarray}%
\nu_l(\bar{\nu_l}) + N(e^-) & \rightarrow l^\mp(e^\mp) + X,%
\quad \quad & (CC) \label{CC} \\%
\nu_l(\bar{\nu_l}) + N(e^-) & %
\rightarrow \nu_l(\bar{\nu_l}) + X, %
\quad \quad &(NC) \label{NC} %
\end{eqnarray}%
where $l=e,\mu,\tau$. These processes also modify the observed
neutrino spectrum; neutrinos are both absorbed in (\ref{CC}) and
driven to lower energies in (\ref{NC}) on their way from a source
to a detector \cite{BGZR}.

Cross-sections of $\nu e$-scattering are very small as compared
with $\nu N$-cross-sections. An important exclusion is the
resonant $\bar{\nu}_e e^-$-scattering \cite{BGRes}:
\begin{equation}%
\bar{\nu}_e + e^- \rightarrow W^- \rightarrow q_i +\bar{q}_j%
\rightarrow \mbox{\emph{hadrons}}. %
\label{res}%
\end{equation}%
The resonance energy of neutrino is $E_0 = m_W^2/2m_e \simeq 6.4
\times 10^{15}$~eV; the hadrons are produced as a spike with
energy $E_h=E_0$. The number of resonant events in the underground
detector with the number of electrons $N_e$ is given by the simple
formula \cite{BGRes}:
\begin{equation}%
\nu_{res}= 2\pi N_e \sigma_{eff}E_0 I_{\bar{\nu}_e}(E_0), %
\label{nures}%
\end{equation}%
where $\sigma_{eff} = (3\pi/\sqrt{2})G_F=3.0\times
10^{-32}$~cm$^2$ is the effective cross-section ($G_F$ is the
Fermi constant) and $2\pi$ is the solid angle open for a deep
underground detector (within another $2\pi$ angle neutrinos are
absorbed).

A detailed discussion of all above mentioned processes can be
found in Ref.~\cite{Gandhi}.

As regards $\nu N$-cross-sections, especially at extremely high
energies, they are unknown yet. Really, to calculate the rate of
high-energy events in a neutrino detector one actually needs the
differential cross-sections of $\nu(\bar\nu) N$ \emph{DIS} in the
whole range of kinematic variables $0 \leq x \leq 1$ and $0 \leq
Q^2 \leq \infty$. Such cross-sections, with \emph{QCD}-effects
being taken into account (see Ref.~\cite{BGnucl}), may be
expressed in terms of \emph{Parton Distribution Functions (PDF)}
in proton, in our case quarks, $q_i(x,Q^2)$, where
$q_i=u,d,c,s,t,b$. However, an influence of non-perturbative
\emph{QCD}-effects on nucleon \emph{Structure Functions (SF)} at
small both $Q^2$ and $x$ cannot be accurately estimated. It makes
one to rely just on theoretical models, i.e.\ on various
extrapolations.

In fact, differential neutrino-nucleon cross-section can be
parameterized with the help of two structure functions,
$F_{2}^{\nu N}( x,Q^2)$ and $F_{3}^{\nu N,\bar\nu N}( x,Q^2)$ (see
e.g.\ Ref.'s~\cite{deGroot,FMR,Hill}). In the case of CC
scattering (\ref{CC}) these cross-sections are
\begin{equation}%
\label{sigf2}%
\frac{d^{2}\sigma^{\nu N, %
\bar\nu N}_{CC}(E_\nu,x,y)}{dx dy} = \frac{\sigma_0 S_W}{2}%
\frac{(1 - y + \frac{y^2}{2}) F_2^{\nu N}( x,Q^2) \pm%
(y-\frac{y^2}{2}) xF_3^{\nu N, %
\bar\nu N}(x,Q^2)}{(1+ S_W x y)^{2}},%
\end{equation}%
where $\sigma_0 = \frac{G_F^2 m_W^2}{\pi}$, $y=\frac{E_h}{E_\nu}$,
$S_W = \frac {S}{m_W^2}$ and $S = 2 m_N E_\nu$, the $\pm$ sign
corresponds to $\nu /\bar\nu$ cross-sections, respectively. It is
useful to decompose the $F_{3}^{\nu N, \bar \nu N}$ into singlet
and non-singlet parts,
\begin{equation}%
F_{3}^{\nu N, \bar \nu N}(x,Q^2) = %
F_3^{NS}(x,Q^2) \pm F_3^S(x,Q^2).%
\end{equation}%
The \emph{NC} cross-sections (\ref{NC}) can be presented in a
similar way, with $m_W$ replaced by $m_Z$, $S_W$ replaced by $S_Z
= 2 m_N E_\nu/m_Z^2$, $\sigma_0 = G_F^2 m_Z^2/\pi$ and with
structure functions given by
\begin{eqnarray}%
F_2^{\nu N (NC)}(x,Q^2)&=&(\delta _1^2 + \delta _2^2 + \delta_3^2
+ \delta _4^2) F_{2}^{\nu N} + (\delta _2^2 + \delta _4^2
-\delta_1^2 - \delta _3^2)x F_{3}^{S},
\label{F2NCcont}\\%
F_3^{\nu N (NC)}(x,Q^2) &=&(\delta _1^2 + \delta _2^2 - \delta_3^2
- \delta_4^2) x F_{3}^{NS}.
\label{F3NCcont}%
\end{eqnarray}%
The chiral couplings $\delta_i$ are
\begin{equation}%
\delta_1 = \frac{1}{2} - \frac{2}{3} \sin^2 \theta_W, \;\; %
\delta_2 = -\frac{1}{2} + \frac{1}{3} \sin^2 \theta_W, \;\;%
\delta_3 = -\frac{2}{3} \sin^2 \theta_W, \;\;%
\delta_4 = \frac{1}{3} \sin^2 \theta_W; %
\end{equation}%
we use $\sin^2 \theta_W = 0.23117$.

The quark contents of functions $F_{2}^{\nu N}$, $F_{3}^{NS}$ and
$F_{3}^{NS}$ are as follows:
\begin{eqnarray}%
F_{2}^{\nu N}(x,Q^2) & = & x (q + \bar q), \label{qcf2} \\%
F_{3}^{NS}(x,Q^2) &= & q - \bar q , \label{qcf3NS} \\%
F_{3}^{S}(x,Q^2) & = & 2(s-c+b-t), \label{qcf3S}%
\end{eqnarray}%
with $q(x,Q^2) = u +d +s +c +b +t$ and $\bar q(x,Q^2) = \bar u +
\bar d +\bar s +\bar c +\bar b +\bar t$.

Various parameterizations of $q_i(x,Q^2)$, including recent
versions of \emph{CTEQ, MRST, GRV}, may be found in \emph{PDFLIB}
\cite{PDFLIB}. All these parton distributions were obtained as
\emph{LO/NLO} solutions of
\emph{Dokshitzer-Gribov-Lipatov-Atarelli-Parisi (DGLAP)} equations
with low $Q^2$ distributions $q_i(x,Q_0^2)$ taken from
experimental data at $Q_0^2 \approx 1$~GeV$^2$. The calculated
structure functions have been found valid in a wide range of
$(x,Q^2)$ parameter space:
\[
10^{-5} \le x \le 1, \;\;\; 1 \mbox{\textrm{ GeV}}^2 \le Q^2 \le
10^8 \mbox{\textrm{ GeV}}^2.
\]
\emph{DGLAP} approach is based on perturbative physics. The
smallness of the \emph{QCD} coupling constant
$\alpha_s(k_{\perp}^2)<1$ implies $k_{\perp}^2\geq Q_0^2$. The
value of $Q_0 \sim 0.3$~GeV determines the smallest allowed value
of $k_{\perp}$ and can be viewed as mass of parton in \emph{QCD}
cascade. The minimal value of $x$ in perturbative approach is then
determined $x_{min} \sim Q_0^2/S$, where $S=2 m_N E_{\nu}$ for
$\nu N$-scattering.

However, with $E_\nu$ increasing the ever smaller values of $x=
\frac{Q^2}{S y}$ get important in $\nu N$-scattering. It should be
noted, that \emph{HENA} actually promises the deepest insight in
small-$x$ physics. Indeed, the record measurements today by
\emph{HERA} \cite{HERA} relate to $F_2^{ep}(x,Q^2)$ structure
function with $x \gsim 10^{-5}$, while neutrino-nucleon \emph{SF}
with $E_\nu \sim 10^{19}$~eV are sensitive to $x \lsim 10^{-6}
\div 10^{-8}$. As \emph{LO/NLO DGLAP} dynamics evolves in
$Q^2$-direction with $x$ being fixed, it provides no information
about small-$x$ \emph{SF} behavior. Moreover, the applicability of
\emph{DGLAP} approach to small-$x$ physics seems to be
questionable. It was shown in Ref.~\cite{Land} that \emph{Mellin}
transform of \emph{DGLAP} splitting matrix ${\mathbf p}(z)$,
${\mathbf p}(N) = \int_0^1 dz z^N {\mathbf p}(z)$, suffers from
singularities at $N = 0$. These singularities arise in the
perturbative expansion of $p_{qG}$ and $p_{GG}$ in powers of
$\alpha_S$.

Nowadays, the only small-$x$ solution obtained in the framework of
perturbative \emph{QCD} is the \emph{BFKL}-pomeron \cite{BFKL}.
However, the validity of this asymptotic solution is still to be
checked. Though the importance of this approach is widely
recognized, there were lately many criticisms of the pure
\emph{BFKL}-pomeron solution (see e.g.\ Ref.~\cite{Dok} and
references therein).

A certain modification of \emph{BFKL}-pomeron description has been
proposed in Ref.~\cite{KMS}. It includes a unified
\emph{BFKL}/\emph{DGLAP} evolution equation with a special
\emph{'consistency constraint'} imposed on \emph{BFKL} component.
Authors applied this approach to calculations of neutrino-nucleon
cross-sections and checked manifestations of such solution in
\emph{HENA}.

In this paper we concentrate on the different approach to the
small-$x$ physics, which is developing by \emph{A. Donnachie and
P. V. Landshoff}, hereafter \emph{DL}, with coauthors \cite{Land}.
It is based on the Regge theory inspired description of small-$x$
$ep$-structure function, $F_2^{ep}(x,Q^2)$. The authors have
actually made the simplest possible assumption, namely, that
contributions from branch points of the complex $l$-plane at $t=0$
are much weaker than those from poles. This hypothesis gives a
rather good description of data and may be regarded as a guideline
in $F_2^{ep}(x,Q^2)$ small-$x$ extrapolation search. Though,
predicting the power-law growth of cross-sections at high
energies, this approach violates the unitarity.

According to \emph{DL}, $F_2^{ep}(x,Q^2)$ may be written down as a
sum of three factorized terms,
\begin{equation}%
\label{F2DL} F_2^{ep}(x,Q^2) = \sum_{i=0}^2
f_i(Q^2)x^{-\epsilon_i},
\end{equation}%
where $\epsilon_0, \epsilon_1$ and $\epsilon_2$ relate to the
so-called \emph{'hard pomeron'}, \emph{'soft pomeron'} and
$\varrho,\omega,f,a$ exchanges, respectively. Shapes of $f_i(Q^2)$
and values of $\epsilon_i$ are parameters; they are to be chosen
so that to better fit the data. One of the best and the most
convenient for our purposes set of $f_i(Q^2)$, viz.\
\begin{eqnarray}%
\label{hardp}%
 f_0(Q^2) &=& A_0 \left( \frac{Q^2}{Q^2 + a_0}
\right)^{1+\epsilon_0} \left(1 + X \ln \left(1+\frac{Q^2}{Q_0^2}
\right)\right), \\
\label{softp}%
f_1(Q^2) &=& A_1 \left( \frac{Q^2}{Q^2 + a_1}
\right)^{1+\epsilon_1} \frac{1}{1+\sqrt{Q^2/Q_1^2}}, \\
\label{faexch}%
f_2(Q^2) &=& A_2 \left( \frac{Q^2}{Q^2 + a_2}
\right)^{1+\epsilon_2},
\end{eqnarray}%
was proposed in Ref.~\cite{Land}. With
\begin{equation}%
\label{DLpar}
\begin{array}{lllllllll}
\epsilon_0 & = & 0.418 ,& \epsilon_1 & = & 0.0808 , & \epsilon_2
& = & -0.4525 , \\
A_0 & = & 0.0410 ,\;\; & A_1 & = & 0.387 ,\;\; & A_2 & = & 0.0504 , \\
a_0 & = & 7.13 ,\;\; & a_1 & = & 0.684 ,\;\; & a_2 & = & 0.00291 , \\
X & = & 0.485 ,\;\; & Q^2_0 & = & 10.6 ,\;\; & Q^2_1 & = & 48.0 ,
\end{array}
\end{equation}%
it describes $539$ data points with $\chi^2 \approx 1.016$ per
point.

It is worthy of note, that authors of this approach rather prefer
fits with $f_0 \propto Q^{\epsilon_0}$ at high $Q^2$ \cite{Land}.
In spite of slightly higher $\chi^2$, such fits are very
attractive since their \emph{hard pomeron} term looks similar to
the perturbative \emph{BFKL} solution. However, the direct
identification of these two pomerons seems to remain dubious.

In the most recent version \cite{Land_New} of their approach A.
Donnachie and P. V. Landshoff propose a new parameterization of
$F_2^{ep}(x,Q^2)$ data. They have reduced the number of free
parameters by including of the real photon cross-section data,
$\sigma^{\gamma p}$. Authors argue that this new fit gives a good
description of data also for $x>0.001$ and for $Q^2<5000$~GeV$^2$.
With no additional parameters, they describe successfully the
charm \emph{SF}, $F_2^c(x,Q^2)$, as well.

However, in the new fit authors use again the power-law dependence
of coefficient $f_0(Q^2)$ (\ref{hardp}) at high $Q^2$. It works
good at small $Q^2$, which are characteristic for photoproduction
processes, but it does not extend to $Q^2 \sim m_W^2$, which are
typical for $\nu N$-scattering at $S>>m_W^2$. Such power-law
behavior of the \emph{hard} pomeron term cannot be reconciled with
the quasi-logarithmic dependence on $Q^2$ of \emph{DGLAP}
\emph{SF} at high $Q^2$. Therefore we cannot directly apply this
new successful parameterization to the description of $\nu
N$-cross-sections. Maybe it will be possible after a slight
modification of the form of \emph{hard pomeron} coefficient, which
would coincide at small $Q^2$ with the original one. But in this
paper we are to use the previous, logarithmic, parameterization
(\ref{F2DL})-(\ref{DLpar}).

In fact, all descriptions of small-$x$ \emph{SF} are basically
extrapolations with certain merits and demerits. In this paper we
construct one more set of structure functions $F_2^{ep}$,
$F_2^{\nu N}$, $F_3^{S}$ and $F_3^{NS}$. We want them
\begin{enumerate}
\item%
to be defined in the whole range of variables $0 \leq x \leq 1$
and $0 \leq Q^2 \leq \infty$;
\item%
to comprise both perturbative at high-$x$ description, viz.\
\emph{CTEQ5} \cite{CTEQ}, and non-perturbative at low-$x$ pure
\emph{Regge} theory approach of \emph{DL};
\item%
to be smooth over both variables with limited change of first
derivative over $\log x$ in the interpolation zone.
\end{enumerate}
Hereafter we call this approach \emph{DL+CTEQ5}. We hope it is
appropriate for the purposes of \emph{HENA}.

Using \emph{DL+CTEQ5} functions, we calculate the $CC$, $NC$ and
total ($CC+NC$) $\nu(\bar\nu) N$-cross-sections and compare them
with the results of Ref.'s~\cite{Gandhi,KMS}. We also use for
comparison the cross-sections obtained in the framework of trivial
\emph{'logarithmic}' extrapolation, hereafter \emph{Log+CTEQ5},
\begin{equation}
\label{log} F_i^{\nu N,Log+CTEQ5}(x<x_{min},Q^2) = F_i^{\nu
N,CTEQ5}(x_{min},Q^2)\left(\frac{x}{x_{min}}\right)^{\beta_i(Q^2)}, %
\end{equation}
\begin{equation}
\label{beta}
\beta_i(Q^2) = \left. \frac{\partial \ln F_i^{\nu N, CTEQ5}(x,Q^2)}%
{\partial \ln x} \right|_{x=x_{min}} ; \;\;\; %
x_{min} = 1\times 10^{-5},
\end{equation}%
which is analogous to the approach of Ref.~\cite{BGZR}. These
\emph{SF} smoothly shoot to the low-$x$ region from the
\emph{CTEQ5} defined high-$x$ one. Starting values of functions
and of their logarithmic derivatives over $x$ in such
extrapolation are taken at the $x=x_{min}$ boundary of
\emph{CTEQ5}.

\section{Construction of DL+CTEQ5 structure functions}
According to (\ref{sigf2},\ref{F2NCcont},\ref{F3NCcont}),
neutrino-nucleon cross-sections depend just on $F_{2}^{\nu
N}(x,Q^2)$, $F_{3}^S(x,Q^2)$ and $F_{3}^{NS}(x,Q^2)$ \emph{SF},
which quark contents is given by (\ref{qcf2}-\ref{qcf3S}).
However, the best small-$x$ data relate to $F_2^{ep}(x,Q^2)$.
Being just different combinations of quark density distributions,
$q_i(x,Q^2)$, these functions are bound indirectly. So, it is
strongly desirable to make use of this information.

\subsection{United $F_2^{ep}$ structure function}
Let's first obtain the united, smooth over $x$ and $Q^2$,
structure function $F_2^{ep}(x,Q^2)$, with small-$x$ behavior
being in accordance with \emph{DL} description
(\ref{F2DL}-\ref{faexch}) and with large-$x$ one being determined
by \emph{CTEQ5}. These functions are different, of course, so that
we are to meet them smoothly.

\begin{figure}
\includegraphics[width=\textwidth]{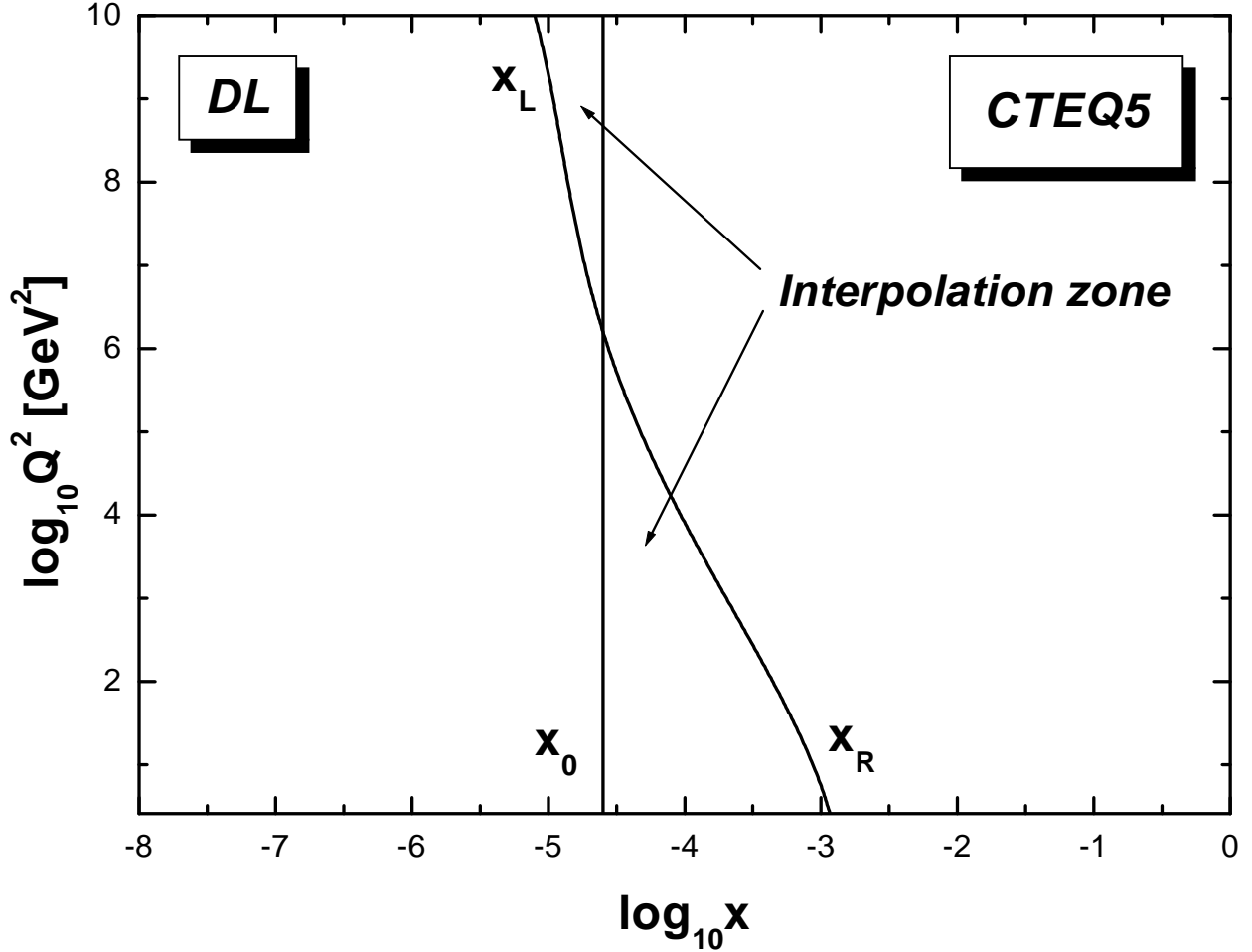}
\caption{The borders of interpolation zone
corresponding to $C=0.12$ (see (\ref{Intrpl})).}
\label{fig:borders}
\end{figure}

First, to restrain the $Q^2$ divergency, we choose the \emph{DL}
parameter set (\ref{hardp}-\ref{DLpar}). Really, at large $Q^2$
only the \emph{hard pomeron} term (\ref{hardp}) survives in
(\ref{F2DL}), so that $F_2^{ep}(x,Q^2) \propto \ln Q^2$. On the
other hand, perturbative dynamics predicts an approximately
logarithmic over $Q^2$ asymptotic growth of \emph{SF} as well. In
fact, the \emph{CTEQ5} $Q^2$-dependence of $F_2^{ep}$ at high
$Q^2$ and $x \approx x_{\min}$ is close to logarithmic, but
different; it rather looks like
\begin{equation}
F_2^{ep}(x,Q^2) \propto (\ln Q^2)^{1+\alpha(x)} \label{F2grow}
\end{equation}
with $|\alpha(x)| \ll 1$. So, despite both functions fit to the
same \emph{HERA} data at $x=x_{\min}$, and therefore should
coincide in a wide range of $Q^2$, they inevitably disperse at
large $Q^2$. To reduce this discrepancy, we solved the equation
$\alpha(x)=0$ numerically; the root is
\begin{equation}%
x = x_0 \simeq 2.527 \times 10^{-5}. \label{x_0}
\end{equation}%
We take the line $x(Q^2)=x_0$ as one of boundaries of the
interpolation zone. It is shown in the Fig.~\ref{fig:borders}.

Luckily, $x_0$ is rather close to the left \emph{CTEQ5} boundary
$x_{min} = 1 \times 10^{-5}$. At $x \simeq x_0$ both \emph{DL} and
\emph{CTEQ5} parameterizations still keep valid. It means that at
relatively small $Q^2 \lsim 500$~GeV$^2$ they are to be very
close, if not equal. It is important, that at large $Q^2$ and
$x=x_0$ both descriptions practically  do not disperse.

On having reconciled the $Q^2$ behaviors, we take concern of
smooth \emph{SF} meeting over $x$ at each $Q^2$. The
$x$-dependencies of these parameterizations are different. To meet
them smoothly, we undertake an interpolation between $\ln
F_2^{DL}$ and $\ln F_2^{CTEQ5}$ with the help of cubic over $\ln
x$ polynomials; these procedure assures the first derivatives to
be continuous in the whole interpolation zone.

So, keeping one border of the interpolation zone at $x(Q^2) = x_0$
and varying  the shape of another border, we call the latter
$x_{L,R}(Q^2)$, one gets a set of different interpolations. The
subscripts $L,R$ mean that at small $Q^2$ we look for the right
border, $x_R(Q^2)$, while at large $Q^2$ we look for the left one,
$x_L(Q^2)$. The crossing takes place at $Q^2 \approx 1.78 \times
10^6$~GeV$^2$. These borders allow to extend the influence of
\emph{DL} description at small $Q^2$ to $x_0<x<x_R$, on the one
hand, and, on the other hand, increase the influence range of the
perturbative description at high $Q^2$ to smaller $x_L<x<x_0$.
Such slant border seems us to be reasonable from the physical
point of view.

The quality of interpolation may be parameterized by imposing of
an additional condition:
\begin{equation}
\max \left| \left. \frac{\partial \ln F_2^{\nu N}(x,Q^2)}{\partial
\ln x} \right|_{x=x^{\prime}} %
- \left. \frac{\partial \ln F_2^{\nu N}(x,Q^2)}{\partial \ln x}
\right|_{x=x^{\prime \prime}} \right| < C, %
\quad \quad%
\forall \left\{x^{\prime}, x^{\prime \prime} \in[x_0,x_{L,R}]
\right\} . \label{Intrpl}
\end{equation}
This actually constrains the maximum change of the tangent in the
$x_0,x_{L,R}$ range. The higher is $C$, the closer borders are
allowed. However, at the same time the higher tangent change
becomes possible in the interpolation zone. An optimum value of
this parameter seems to be $C = 0.12$.

The interpolation zone borders corresponding to this value of $C$
are plotted in the Fig.~\ref{fig:borders}.

\begin{figure}
\includegraphics[width=\textwidth]{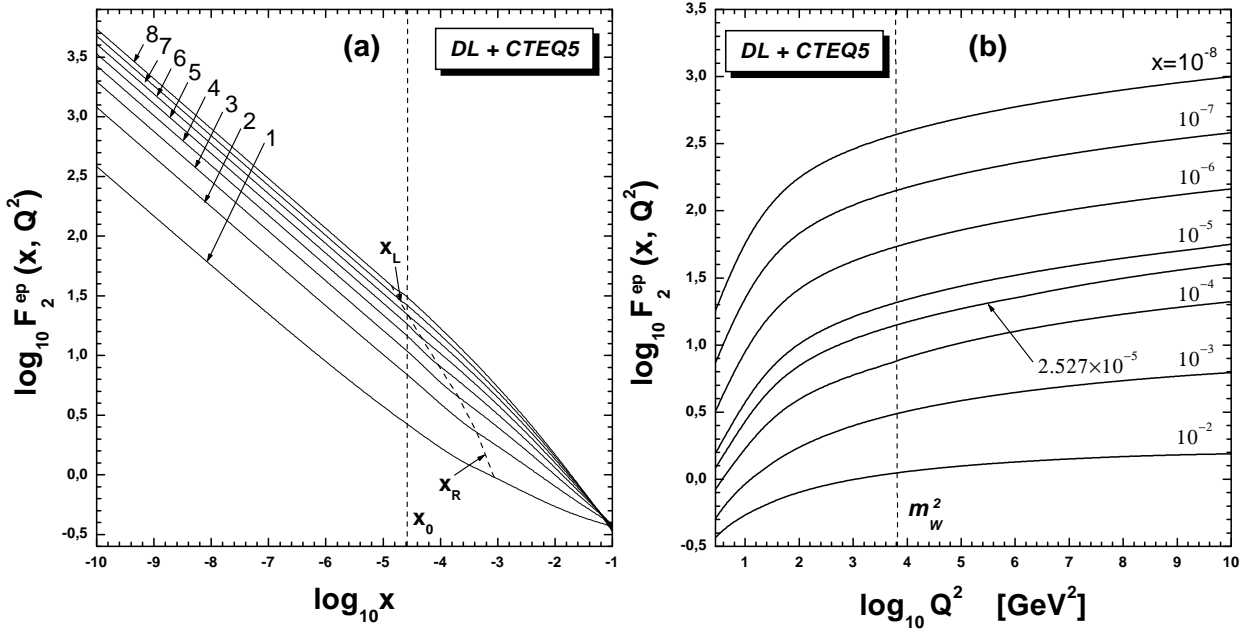}
\caption{a) $F_2^{ep}(x,Q^2)$ as a function of $x$ for different
values of $\log_{10}Q^2 = 1,2,3,4,5,6,7,8$. Each label equals to
the value of corresponding $\log_{10}Q^2$. The cubic spline
interpolation over $x$ zone lies between two dash lines. b)
$F_2^{ep}$ as the function of $Q^2$ for several values of $x$
denoted in the plot. The $m_W$ line corresponds to $Q^2 = m_W^2$.}
\label{fig:F2ep}
\end{figure}
So, the described procedure results in $F_2^{ep}(x,Q^2)$ with the
desired properties. The $x$-dependence of these structure
functions is plotted in Fig.~\ref{fig:F2ep}a for several values of
$Q^2$. In Fig.~\ref{fig:F2ep}b the $Q^2$-dependence of the same
functions is depicted for several values of $x$.

\subsection{United neutrino-nucleon structure functions}
\begin{figure}
\includegraphics[width=\textwidth]{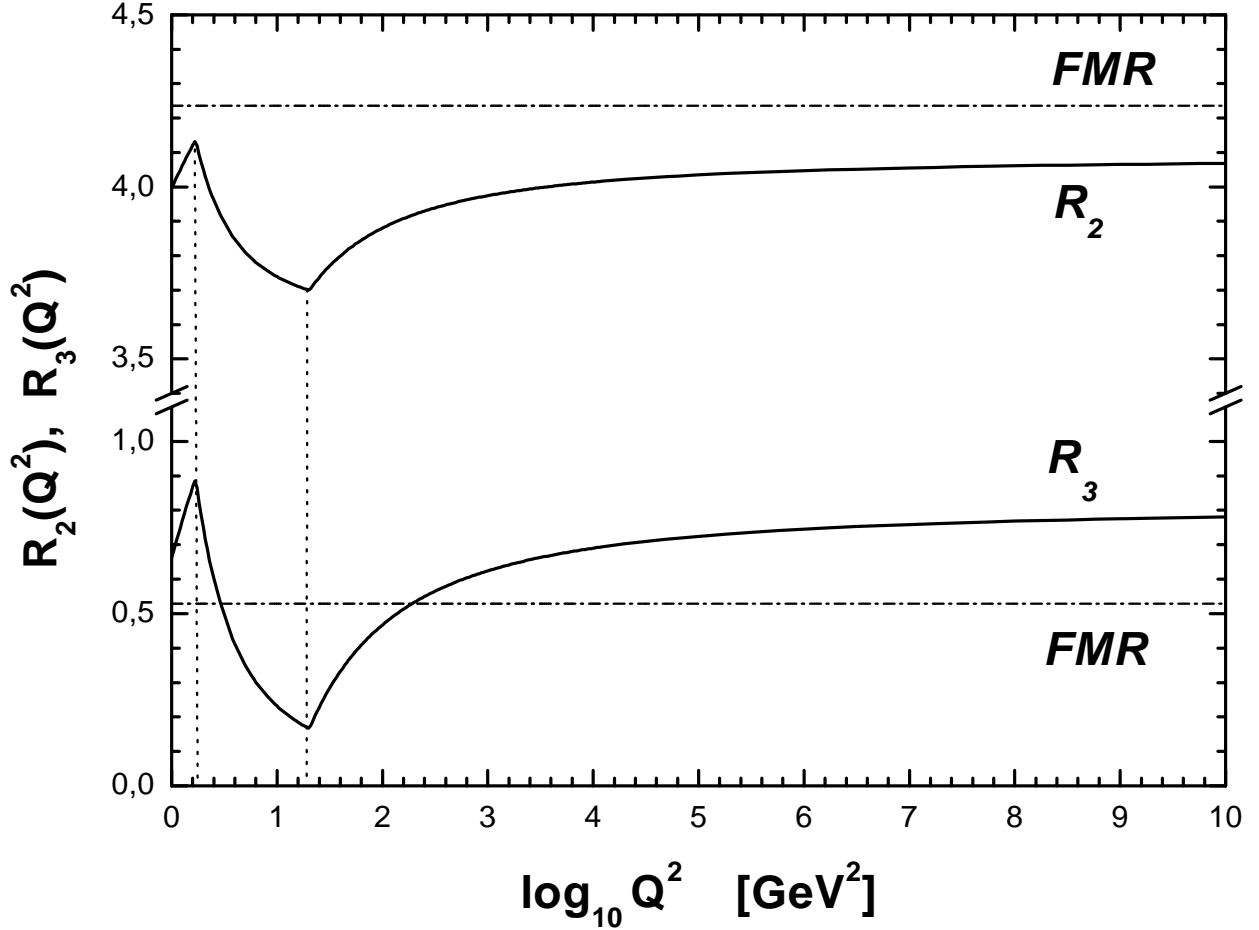}%
\caption{Ratios $R_2(Q^2) = \frac{F_2^{\nu N}}{F_2^{ep}}$ and
$R_3(Q^2) = \frac{x F_3^S}{F_2^{ep}}$ calculated within
\emph{CTEQ5}\ at $x=x_0$. The \emph{FMR} lines correspond to
analogous coefficients $R_2^{FMR} = \frac{72}{17}$ and $R_3^{FMR}
= \frac{9}{17}$.} \label{fig:R23}
\end{figure}

Now let us turn to construction of neutrino-nucleon \emph{SF}
$F_{2}^{\nu N}(x,Q^2)$, $F_{3}^S(x,Q^2)$ and $F_{3}^{NS}(x,Q^2)$.
For relation of $F_2^{ep}(x,Q^2)$ with $F_2^{\nu N}(x,Q^2)$ a
simple receipt has been proposed in Ref.~\cite{FMR}, hereafter
\emph{FMR}. Under an assumption that $i$-th flavor quark and
anti-quark distributions in proton are equal, $q_i = \bar{q}_i$,
and that $ u = d = s = 2c = 2b$, the
\[
F_2^{\nu N, FMR}(x,Q^2) = \frac{72}{17} F_2^{ep}(x,Q^2)
\]
rule had been derived there. To get better description, we modify
this approach by introducing of $Q^2$-dependent ratios $R_2(Q^2)$
and $R_3(Q^2)$. These ratio functions may be extracted from
\emph{CTEQ5} at $x=x_0$ according to the rule
\begin{eqnarray}%
R_2(Q^2) &= &\frac{F_2^{\nu N, CTEQ5} (x_0,Q^2)}{F_2^{ep, CTEQ5}
(x_0,Q^2)}, \label{R2} \\
R_3(Q^2) &= &\frac{x F_3^{S, CTEQ5}(x_0,Q^2)} {F_2^{ep,
CTEQ5}(x_0,Q^2)}; \label{R3}
\end{eqnarray}%
they are plotted in Fig.~\ref{fig:R23}. These ratios differ from
the constant values $R_2=\frac{72}{17}$ and $R_3=\frac{9}{17}$
(this value we derived using the assumption of Ref.~\cite{FMR}),
denoted in the picture as \emph{FMR}. The difference is especially
appreciable at small $Q^2$ due to the thresholds of heavy quarks
production.

Combining ratios (\ref{R2},\ref{R3}) with the constructed
$F_2^{ep}(x,Q^2)$ and assuming that these relations keep valid at
lower values of $x$, we get the small-$x$
\begin{eqnarray}%
F_2^{\nu N}(x,Q^2) &= &R_2(Q^2)\times F_2^{ep}(x,Q^2),
\label{F2NDL} \\%
F_3^{S}(x,Q^2) &= &R_3(Q^2)\times F_2^{ep}(x,Q^2). \label{F3DL}
\end{eqnarray}%
At the next step we undertake the analogous to (\ref{Intrpl})
cubic polynomial interpolation. It allows to smooth and restrict
the $x$-discontinuity of these parameterizations.

To describe the negligible small-$x$ non-singlet structure
function $F_3^{NC}(x,Q^2)$, we use a trivial extrapolation of the
corresponding \emph{CTEQ5} function in a way analogous to the
\emph{Log+CTEQ5} (\ref{log},\ref{beta}). This completes the
construction of united neutrino-nucleon \emph{SF}.

\begin{figure}
\includegraphics[width=\textwidth]{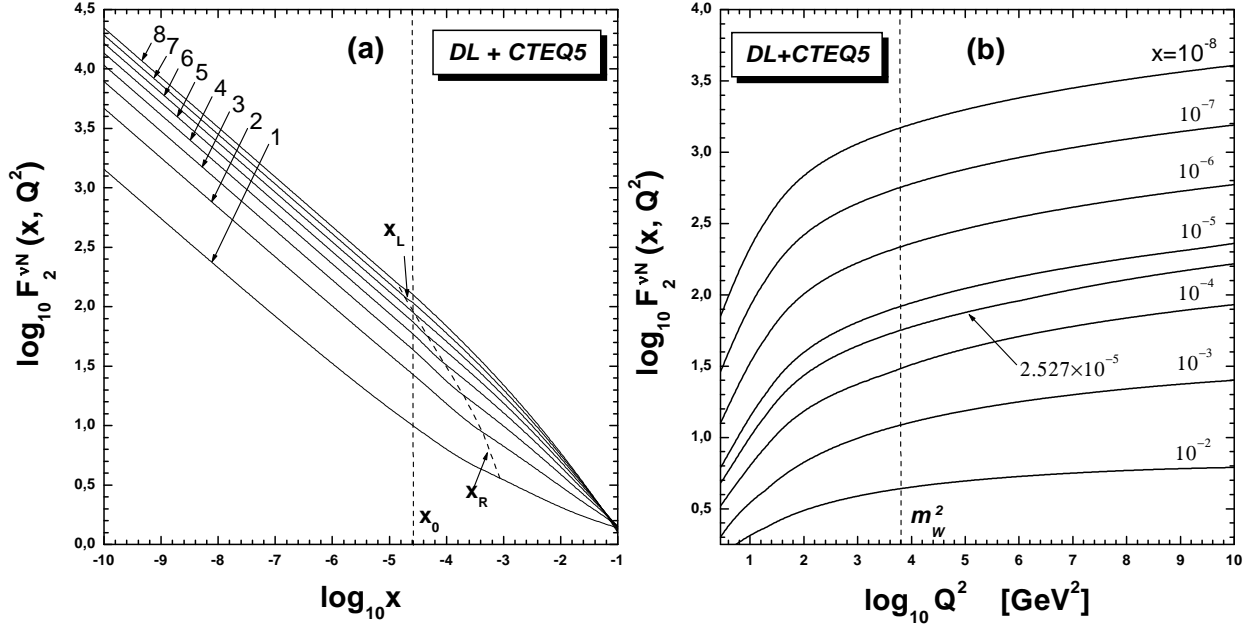}
\caption{a) $F_2^{\nu N}(x,Q^2)$ as a function of $x$ for
different values of $\log_{10}Q^2 = 1,2,3,4,5,6,7,8$. Each label
equals to the value of corresponding $\log_{10}Q^2$. The cubic
spline interpolation over $x$ zone lies between two dash lines. b)
$F_2^{\nu N}$ as the function of $Q^2$ for several values of $x$
denoted in the plot. The $m_W$ line corresponds to $Q^2 = m_W^2$.}
\label{fig:F2meet}
\end{figure}
The characteristic features of the derived \emph{SF} are
illustrated in Fig.'s~\ref{fig:F2meet}a and \ref{fig:F2meet}b.
$F_2^{\nu N}(x,Q^2)$ are depicted there for several values of $x$
versus $Q^2$ and for several values of $Q^2$ versus $x$.

\begin{figure}
\includegraphics[width=\textwidth]{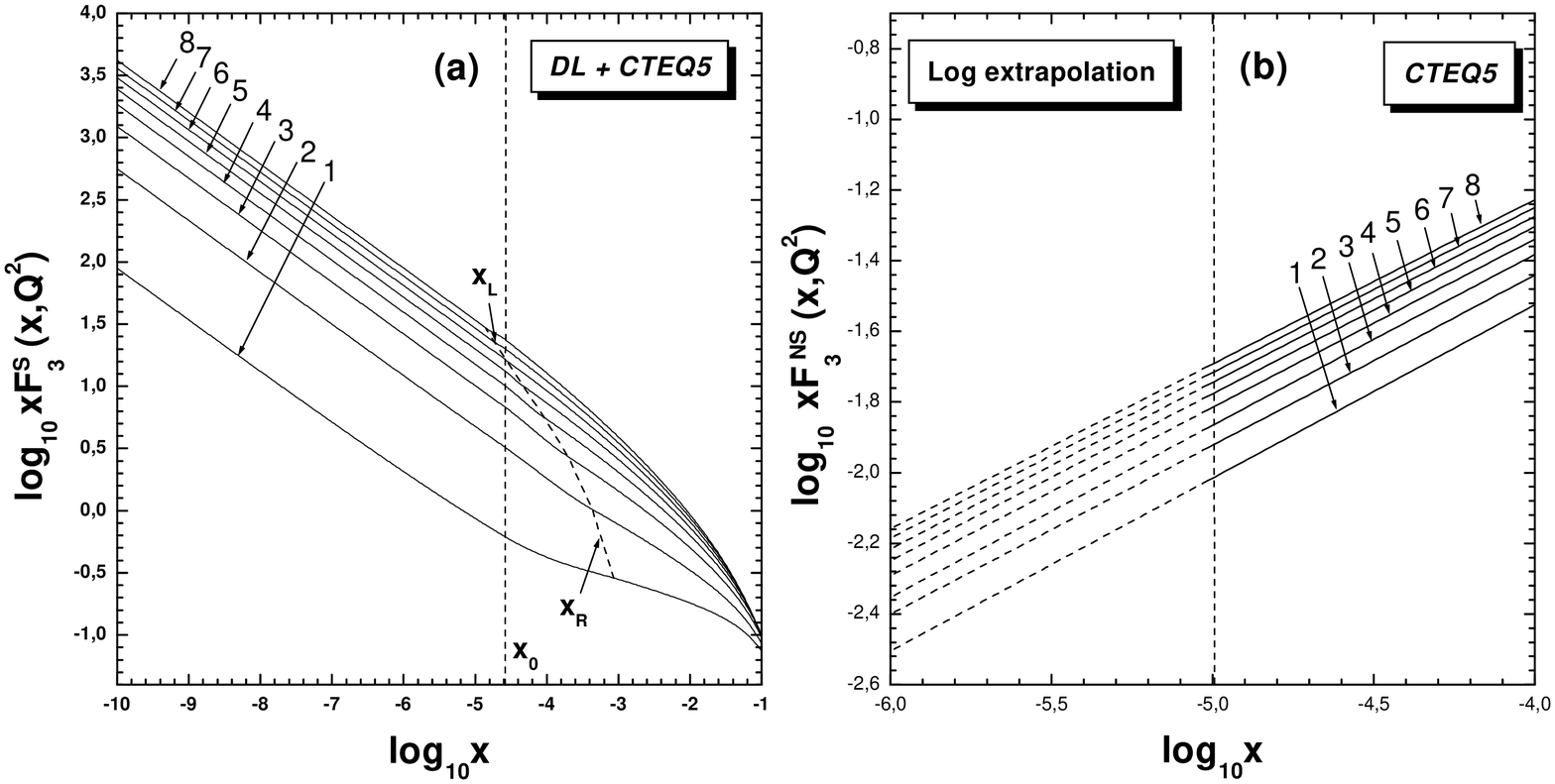}
\caption{a) $x F_3^{S}(x,Q^2)$ as a function of $x$ for different
values of $\log_{10}Q^2 = 1,2,3,4,5,6,7,8$. Each label equals to
the value of corresponding $\log_{10}Q^2$. The cubic spline
interpolation over $x$ zone lies between two dash lines. b) The
same plot as a) but for non-singlet $x F_3^{NS}(x,Q^2)$. The
dashed lines correspond to logarithmic extrapolation of
corresponding perturbative function as described in the text.}
\label{fig:F3meet}
\end{figure}
The behavior of $x F_3^{S}(x,Q^2)$ and of $x F_3^{NS}(x,Q^2)$ as
functions of $x$ are demonstrated in Fig.'s~\ref{fig:F3meet}a and
\ref{fig:F3meet}b.

\section{Comparison of cross-sections}
Substituting the constructed \emph{DL+CTEQ5} \emph{SF} into
Eq.'s~(\ref{sigf2},\ref{F2NCcont},\ref{F3NCcont}), we obtain the
differential $\nu N$-cross-sections. The following integration
over $x$ and $y$ and summation of \emph{CC}- and \emph{NC}-inputs
gives the total cross-sections as functions of $E_\nu$.

\begin{figure}
\includegraphics[width=\textwidth]{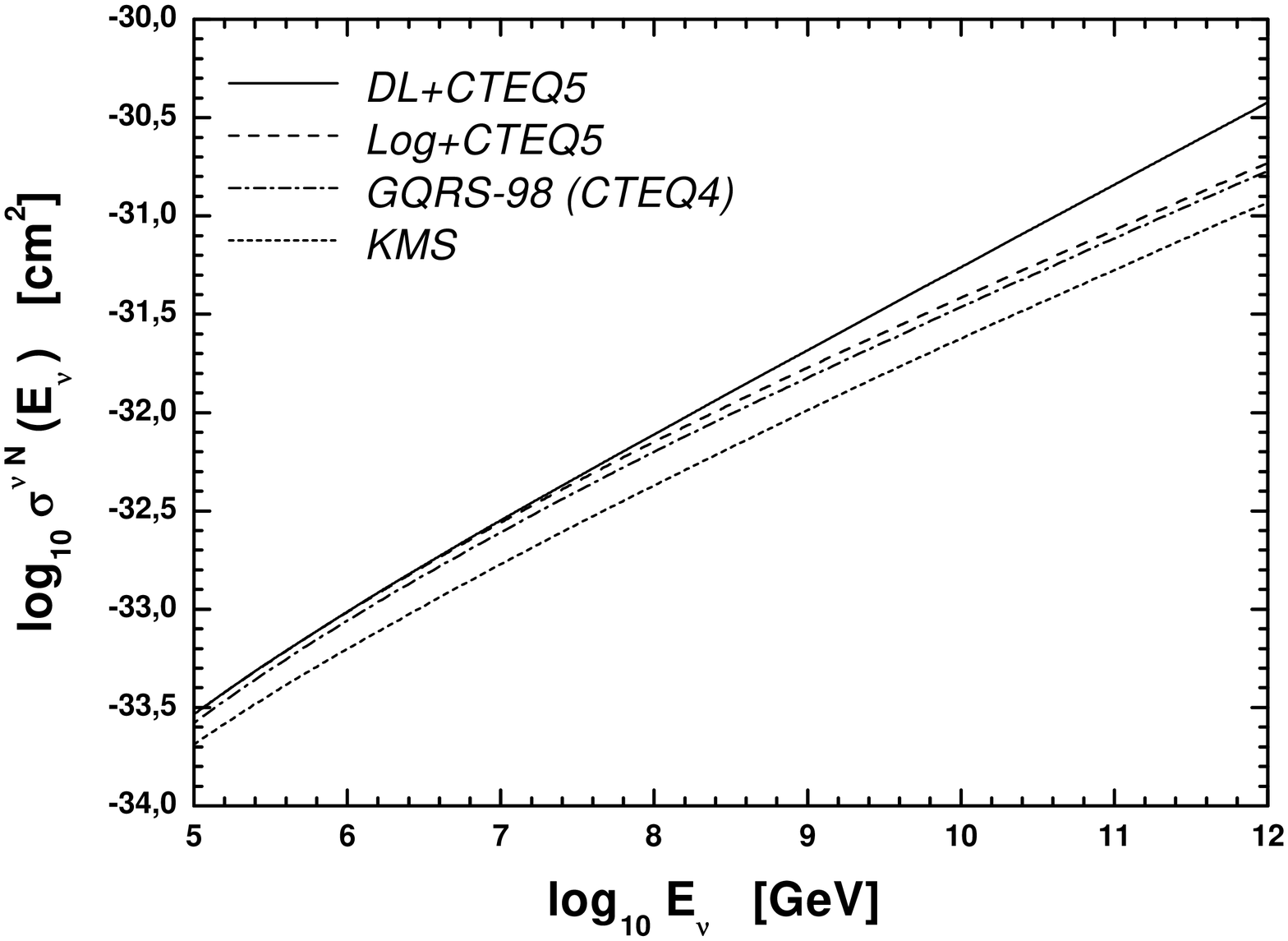}
\caption{ Total \emph{(CC+NC)} \emph{DLGLAP+CTEQ5} $\nu
N$-cross-sections. The analogous cross-sections of $Log +
\emph{CTEQ5}$ extrapolation (\ref{log}), of \emph{CTEQ4}
parameterization by \emph{Gandhi et al.} and of the united
\emph{BFKL/DGLAP} approach by \emph{Kwiecinski, Martin and
Stasto}, labelled as \emph{KMS} are shown for comparison.}
\label{fig:crsscn}
\end{figure}
Denoted as \emph{DL+CTEQ5}, this sum is shown in
Fig.~\ref{fig:crsscn}. For comparison we also plot here the
corresponding cross-sections obtained in the framework of
\begin{description}
 \item[a)] simple $Log + \emph{CTEQ5}$ extrapolation (\ref{log});
 \item[b)] \emph{CTEQ4} parameterization by \emph{Gandhi et al.}
 Ref.~\cite{Gandhi}, denoted as \emph{GQRS-98 (CTEQ4)};
 \item[c)] the united \emph{BFKL/DGLAP} approach by \emph{Kwiecinski,
 Martin and Stasto} \cite{KMS},
labelled as \emph{KMS}.
\end{description}

\begin{figure}
\includegraphics[width=\textwidth]{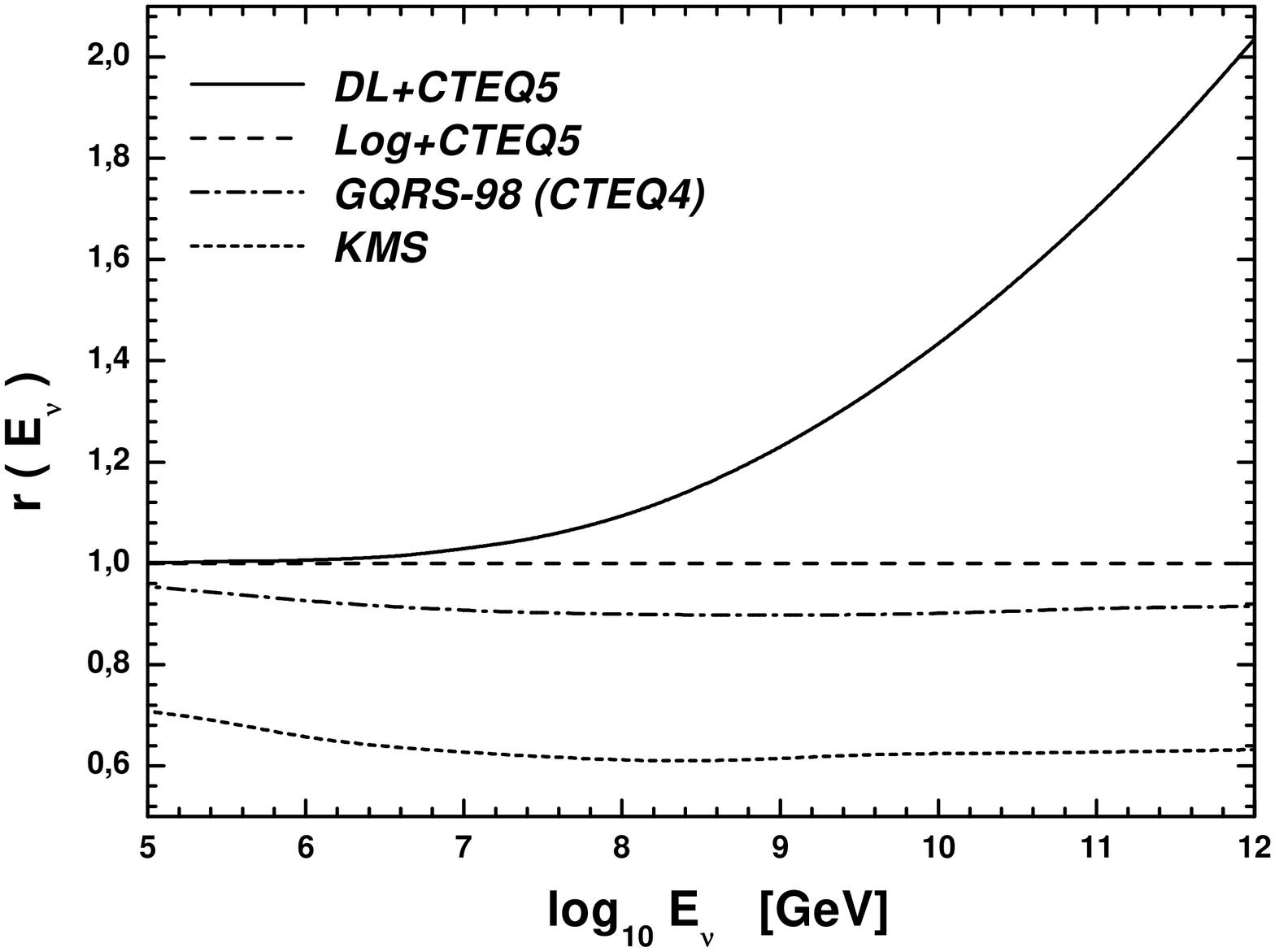}
\caption{$r(E_\nu)$ are ratios of the shown cross-sections and of
\emph{Log}+\emph{CTEQ5} cross-section.} \label{fig:ratios}
\end{figure}
Due to Regge \emph{hard pomeron} pole, our approach predicts the
more rapid growth of cross-sections at high energies. The
differences between these calculations become especially clear in
Fig.~\ref{fig:ratios}. We have divided each cross-section by
corresponding cross-section of \emph{Log}+\emph{CTEQ5}. The ratios
are plotted in the graph. At $E_\nu = 1 \times 10^{12}$~GeV the
\emph{DL}+\emph{CTEQ5} turns out to be twice as high as the
\emph{logarithmic} cross-section.

This difference is neither unexpected nor dramatic for
\emph{HENA}. One should remember that uncertainties in
$\nu$-fluxes are much higher, while expected low measurement
accuracy of future detectors and scarce statistics suggest, that
such difference may be practically insignificant. However, we
believe that rapid growth of $\nu N$-cross-sections may be
eventually discovered in future giant detectors. This effect may
play the essential role for \emph{UHE} $\nu^{\,\prime}s$
predicted in the framework of \emph{TD} models.

One should also keep in mind, that \emph{DL} Regge theory approach
violates unitarity. It implies that the predicted power-law growth
of $\nu N$-cross-sections should be replaced at higher energies
by, say, $\sigma \propto \ln^2E_\nu$ one. Though it is yet unknown
where and how this occurs.

\section*{Conclusions}
In this paper we have derived a new full set of smooth over $x$
and $Q^2$ $ep$- and $\nu N$-structure functions, which are defined
for arbitrary allowable values of these kinematic variables.
According to construction, these functions are in agreement with
Regge theory inspired \emph{hard + soft pomeron} small-$x$
parameterization by Donnachie and Landshoff, and coincide with
perturbative $QCD$ parameterization by \emph{CTEQ5} at large $x$.
For the smooth meeting of these structure functions over both $x$
and $Q^2$, the special interpolation zone boundaries have been
defined.

We recalculate the known $F_2^{ep}(x,Q^2)$ to $F_{2,3}^{\nu N
}(x,Q^2)$ structure functions at small $x$'s with the help of
introduced ratios $R_{2,3}(Q^2)$, which are derived from
perturbative \emph{CTEQ5} description at $x_0=2.527 \times
10^{-5}$.

Using these new structure functions, we have calculated the $\nu
N$-cross-sections at extremely high energies and compared them
with those earlier obtained a) within a simple logarithmic
extrapolation of perturbative structure functions
(\emph{Log+CTEQ5}) and b) in papers \cite{Gandhi,KMS}. At small
and moderate energies these cross-sections are practically
indistinguishable. However, at extremely high energies
non-perturbative \emph{hard pomeron} dynamics causes a quicker
rise of total $\nu N$-cross-sections with energy. Actually, these
growth is the highest among all ever predicted in the framework of
conventional theories.

We understand that pure pomeron behavior of \emph{SF} cannot be a
final answer since it violates unitarity. Nevertheless, we believe
that such approach may be relevant in a wide range of energies
involved in \emph{HENA}.

\section*{Acknowledgments}
This work was supported by the \emph{INTAS} grant No: 99-1065. We
are grateful to V. S. Berezinsky for many advice and cooperation
in this work. A.~G. thanks M.~S. Sergeenko and G.~Navarra for
stimulating discussions. We are also grateful to the unknown
referee for the attraction of our attention to the most recent
data and to the paper Ref.\cite{Land_New}.

%%%%%%%%%%%%%%%%%%%%%%%%%%%%%%%%%%%%%%%%%%%%%%%%%%%%%%%%%%%%%%%%

\end{document}